\documentclass[twocolumn,aps,prl,showpacs,amsmath,superscriptaddress,longbibliography,notitlepage]{revtex4-1}
\usepackage{amssymb}
\usepackage{mathrsfs}
\usepackage{graphicx}
\usepackage{float}
\usepackage[caption=false]{subfig}
\usepackage[pdftex,colorlinks=red]{hyperref}
\usepackage{verbatim}

\usepackage{epstopdf}

\usepackage[normalem]{ulem}
\usepackage{verbatim}
\usepackage{xcolor}
\usepackage{bm}

\def\red{\textcolor{red}}
\def\blue{\textcolor{blue}}
\def\magenta{\textcolor{magenta}}

\def\GG{\textcolor{black}}

\begin{document}

\title{Quantized classical response from spectral winding topology}


\author{Linhu Li}  \email{lilh56@mail.sysu.edu.cn}
\affiliation{Guangdong Provincial Key Laboratory of Quantum Metrology and Sensing $\&$ School of Physics and Astronomy, Sun Yat-Sen University (Zhuhai Campus), Zhuhai 519082, China}
\author{Sen Mu} 
\affiliation{Department of Physics, National University of Singapore, Singapore 117551, Republic of Singapore}
\author{Ching Hua Lee} \email{phylch@nus.edu.sg}
\affiliation{Department of Physics, National University of Singapore, Singapore 117551, Republic of Singapore}
\author{Jiangbin Gong}  \email{phygj@nus.edu.sg}
\affiliation{Department of Physics, National University of Singapore, Singapore 117551, Republic of Singapore}
\begin{abstract}
Topologically quantized response is one of the focal points of contemporary condensed matter physics. While it directly results in quantized response coefficients in quantum systems, there has been no notion of quantized response in classical systems thus far. {This is because quantized response has always been connected to topology via linear response theory that assumes a quantum mechanical ground state. Yet, classical systems can carry arbitrarily amounts of energy in each mode, even while possessing the same number of measurable edge modes as their topological winding.} In this work, we discover the totally new paradigm of \emph{quantized classical response}, which is based on the spectral winding number in the complex spectral plane, rather than the winding of eigenstates in momentum space. 
Such quantized response is classical insofar as it applies to phenomenological non-Hermitian setting, arises from fundamental mathematical properties of the Green's function, and shows up in steady-state response, without invoking a conventional linear response theory. Specifically, the ratio of the change in one quantity depicting signal amplification to the variation in \GG{one imaginary flux-like parameter} 
is found to display fascinating plateaus, with their quantized values given by the spectral winding numbers as the topological invariants. 

\end{abstract}

\date{\today}

\maketitle
\section{Introduction}

Topological quantization has captivated a generation of physicists since the discovery of the quantum Hall effect \cite{
QHE}. In the more recent years, its standing as a novel quantum phenomenon was further strengthened by the observation of discretely varying Hall conductances in the quantum anomalous Hall \cite{Haldane_QAHE,liu2016quantum} and quantum spin Hall effects \cite{KM_QSHE1}. 
Rigorously established through linear response theory, the link between topological winding numbers and quantized conductivity has indeed earned a place as a classic result of quantum condensed matter physics. Indeed quantized Hall response and nontrivial Chern topology are now widely regarded to be almost synonymous.

Yet, the concept of quantized response has so far eluded classical systems. Without a quantum mechanical ground state, classical systems are not amendable to conventional linear response theory, which expresses quantized responses as perturbations upon a ground state. While classical metamaterials like photonic crystals, acoustic structures and electrical circuits can possess an integer number of topologically protected boundary modes~\cite{huber2016topological,nash2015topological,lu2014topological,imhof2018topolectrical}, their response behaviors are not
based on the number of accessible channels, but on analog solutions of differential equations that are by no means quantized. 

In this work, we introduce the paradigm of quantized response based on the winding of the spectrum in the complex energy plane, rather than the winding of eigenstates in momentum space.
For a long time, the complex spectral winding number has been exploited in predicting the non-Hermitian pumping under the open boundary conditions (OBCs) [known as the non-Hermitian skin effect (NHSE)] \cite{yao2018edge,okuma2020topological,zhang2019correspondence,Lee2019anatomy}, leading to the breaking of bulk-boundary correspondence and various anomalous topological phenomena
\cite{xiong2018does,shen2018topological,
kawabata2019symmetry,
yao2018edge,
Lee2019anatomy,
kunst2018biorthogonal,Yao2018nonH2D,Yin2018nonHermitian,Hui2018nonH,li2019geometric,song2019realspace,song2019non,okuma2019topological,mu2020emergent,
jiang2019interplay,longhi2019topological,
Song2019BBC,
Lee2019hybrid,li2020topological,
helbig2020generalized,xiao2020non,
okuma2020topological,zhang2019correspondence,
lee2020unraveling,li2020critical,yang2019auxiliary,yi2020nonH
}. However, to date no directly measurable quantity has been associated with the spectral winding. Notably, spectral winding as a topological feature is not directly related to quantum physics and is hence a \emph{classical} concept. Furthermore, the notion of classical response can be also a property of the Green's function alone. While nontrivial complex energy winding is in principle well-defined for quantum systems, it is most physically relevant in classical settings like mechanical, photonic, plasmonic and electrical systems where non-Hermiticity does not present significant measurement difficulties~\cite{ghatak2020observation,xiao2020non,helbig2020generalized}. Indeed, electrical circuits are governed by circuit Laplacians whose complex eigenvalues merely indicate phase shifts or steady-state impedances, rather than ephemeral excitations.


{Specifically, what we find quantized is the response of the logarithm of the Green's function components with respect to an imaginary flux-like local parameter that continuously adjust the system boundary conditions, forming some quantum-Hall-like plateaus quantized according to the spectral winding number. This discovery was inspired by the observation that deforming a non-Hermitian system from periodic to open boundaries (PBCs to OBCs)~\cite{xiong2018does,Lee2019anatomy,Lee2019hybrid,lee2020unraveling,mu2020emergent}, which is closely related to complex flux insertion, always reduces the spectral winding number one by one till it reaches zero.} In a classical setup subject to a steady-state drive 
{e.g. a circuit lattice with an input current}, the quantized quantity \GG{can be}  the logarithmic impedance experienced by the response field, e.g. the voltage. This intriguing result is rooted in the way non-Bloch eigenstates explore the interior of spectral loops, and can \GG{be understood with an unexpected type of non-Hermitian pumping that is scale-free} 
~\cite{li2020impurity}. 

\section{Results}
\subsection{Classical vs. Quantum response}

{In conventional literature, quantized response usually refers to quantized linear response in a quantum setting, where an occupied quantum state is driven by a time-dependent perturbation. In this work, however, our focus is to highlight that it is still possible to extract a quantized quantity from the response of a purely classical system, where there is no notion of an occupied ground state. In the following, we shall first clarify the distinction between quantum and classical response.}

{For concreteness, first consider the quantum setting described by a Hamiltonian $\hat H(t)=\hat H_0 + \epsilon(t)\hat H'$, where $\hat H_0$ is the equilibrium Hamiltonian, $\hat H'$ the perturbed operator and $\epsilon(t)$ a time-dependent factor that controls the perturbation strength. The linear response of a chosen operator $\hat \phi$ (which has zero expectation when $\epsilon(t)=0$) is its expectation $\hat \phi(t)$ after $\epsilon(t)$ is switched on. In frequency space, we write
\begin{equation}
\langle \hat \phi\rangle(\omega)=D(\omega)\epsilon(\omega),
\end{equation}
where
\begin{eqnarray}
D(\omega)& =& -i\int_0^\infty \langle[\hat \phi(t),\hat H'(0)]\rangle \,dt\notag\\
& =& -iZ^{-1}\int_0^\infty \text{Tr}\left(e^{-\beta H_0}[\hat \phi(t),\hat H'(0)]\right) \,dt\notag\\
&=& \frac1{Z}\sum_{nm}\frac{\langle n|\hat \phi|m\rangle\langle m|\hat H'|n\rangle (e^{-\beta E_n}-e^{-\beta E_m})}{\omega +i\delta -(E_m-E_n)}
\label{Kubo}
\end{eqnarray}
where $Z=\text{Tr}\left(e^{-\beta \hat H_0}\right)$, $\beta $ is the inverse temperature and the trace is taken over all eigenstates $|n\rangle$ satisfying $\hat H_0|n\rangle = E_n|n\rangle$. Very importantly, it has been assumed that the system is in a ground state where the eigenstates $|n\rangle$ are occupied according to the Boltzmann probability distribution $\propto e^{-\beta E_n}$. 
{By expressing the trace as a momentum-space integral which shall correspond to a topological invariant,} 
a topologically quantized response coefficient
can be obtained, for instance, with the DC limit of $D(\omega)/i\omega$ in the case of Hall response.
}

{
By contrast, in the classical settings that we shall focus on i.e. photonic waveguides, acoustic lattices and electronic circuit, the system do not settle into a ground state, and Eq.~\ref{Kubo} is inapplicable. 
The classical response corresponds to the flow of arbitrary amounts of optical, phononic or electric current, and not the modified (quantized) occupancy of quantum mechanical eigenstates.} Consider an external coherent drive $\vec{\epsilon}=(\epsilon_1,\epsilon_2,...\epsilon_L)$ with different amplitudes applied to each of the $L$ sites of a lattice \cite{xue2020non}.  For a {steady-state} drive with a fixed frequency $\omega$, $\vec{\epsilon}(t)=\vec{\epsilon}(\omega){\rm exp}(-i\omega t)$, and the resultant classical response field at the same frequency can be written as $\vec{\phi}(t)=\vec{\phi}(\omega){\rm exp}(-i\omega t)$ with the response field amplitude given by
\begin{eqnarray}
\vec{\phi}(\omega)=G(\omega,\gamma)\vec{\epsilon}(\omega),~G(\omega,\gamma)=\frac{1}{\omega+i\gamma-H},\label{eq:green}
\end{eqnarray}
analogous to Eq.~\ref{Kubo} which is exclusively for quantum settings. Here $G$ is the Green's function matrix and $\gamma$ represents an overall gain/loss in the system.  
For a signal entering the system from a single site $x$, $\vec{\epsilon}$ only possesses one nonzero component $\epsilon_x$, and the induced field at another site $y$ is $\phi_{y}=G_{yx}\epsilon_x$. {In particular, the directional signal amplification of a signal entering one end of a 1D chain and measured at the other end is described by the two matrix elements $G_{1N}$ and $G_{N1}$ \cite{Wanjura2020,xue2020non}}. 

\subsection{Motivation for quantized Green's function response}

{
In the spectral representation, the Green's function Eq.~\ref{eq:green} takes the form
\begin{eqnarray}
G=\sum_n\frac{1}{\omega+i\gamma-E_n}|\Psi_n^R\rangle\langle\Psi^L_n|,\label{eq:green_sr}
\label{G}
\end{eqnarray}
where $|\Psi^{L/R}_n\rangle$ are the left/right eigenstates corresponding to the $n$-th eigenenergy $E_n$. Its matrix elements $G_{xy}$ 
can be computed by evaluating $|\Psi^{L/R}_n\rangle$ at $x$ and $y$. Eq.~\ref{G} is valid in both classical and quantum settings, since the Hamiltonian is just the operator that describes time evolution, and is well-defined regardless of whether position and momentum commute. }

{Ordinarily, we do not expect the matrix elements of $G$ (or functions of them) to respond to an external influence $\beta$ in a quantized manner, since there is no reason why the derivatives of $(\omega-E_n)^{-1}$ and the eigenstates should conspire to add up to an integer. However, when translation symmetry is broken, the eigenstates can potentially become exponentially localized like $\sim e^{\kappa x}$, such that the matrix elements of $G$ are dominated by the largest spatial decay rate $\kappa=\kappa_\text{max}$, with $G_{xy} \sim e^{\kappa_\text{max}(y-x)}$. In such special scenarios, the response of $\ln G_{xy}$ for a fixed interval $x-y$ is \emph{wholly} dependent on how $\kappa_\text{max}$ varies with the external influence. In particular, if $\kappa_\text{max}$ were to vary with an external parameter in a quantized manner, so will the response quantity $\ln G_{xy}$. }

In this work, we discover that $\ln G_{xy}$ indeed possess such a quantized response if the external influence $\beta$ were to be 
an impurity parameter tuning the boundary conditions, which coincides with tuning an imaginary flux when the latter is sufficiently weak \cite{Lee2019anatomy}. 
This quantized quantity is furthermore equal to the winding number of the energy spectrum in the complex energy plane. 
In the following sections, we shall elaborate on these rather surprising findings, and show the classical quantized response can be measured. While this quantization applies to both classical and quantum systems, we shall call it the quantized \emph{classical} response to distinguish it from the topological Hall response that exclusively exists in quantum systems.

\subsection{Point-gap topology and PBC-OBC spectral evolution}
\begin{figure}
\includegraphics[width=1\linewidth]{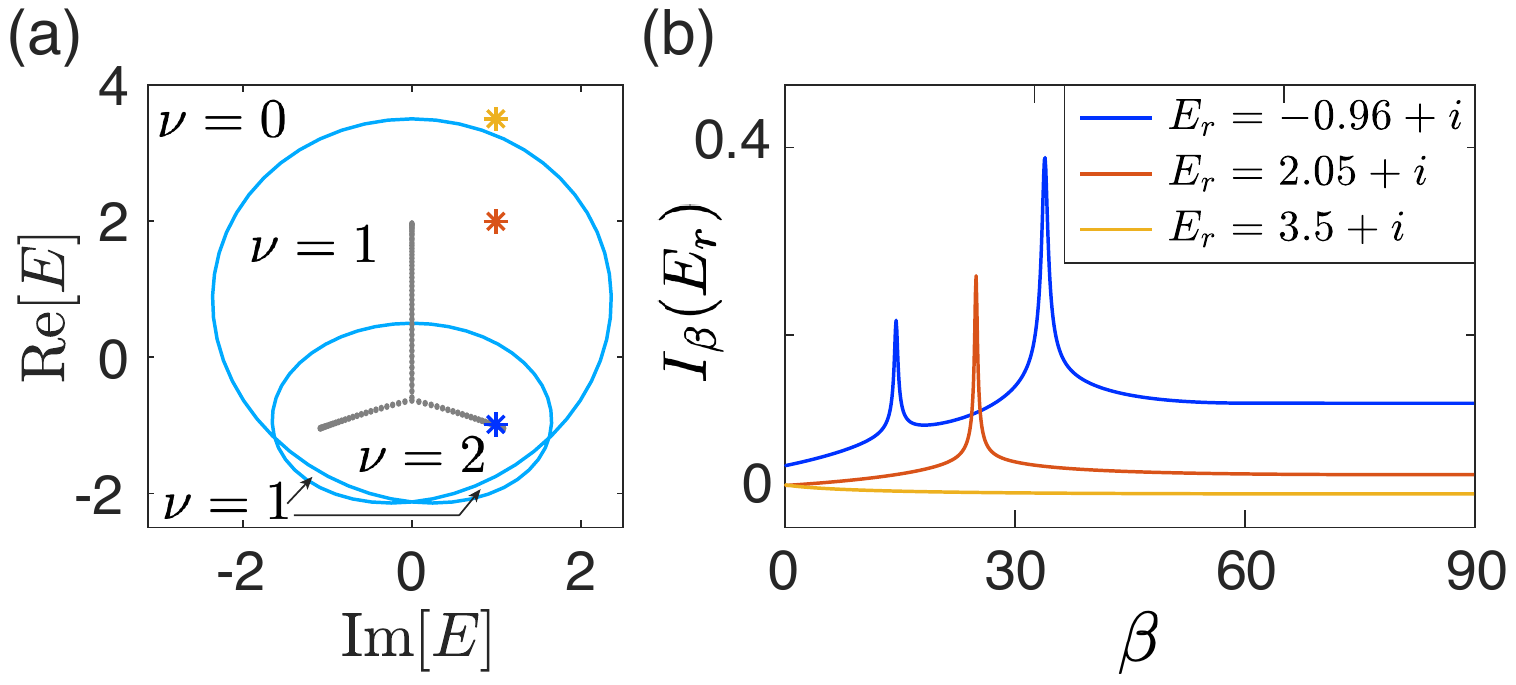}
\caption{
(a) PBC (cyan loops) and OBC (gray dots) spectra of the Hamiltonian in Eq.~ (\ref{eq:H_general}) with $r=l=2$. The spectral winding number for each different regime is indicated on the panel. 
(b) {The quantity $I_\beta(E_r)$ (Eq.~\ref{Ibeta}) vs $\beta$ for the three cases marked in (a). It diverges exactly $\nu$ times when $E_r$ sits in a region of spectral winding $\nu$.}
Parameters are set at $t_1=1$, $t_{-1}=0.5$, $t_2=2$, and $t_{-2}=0$, with $N=100$.
}
\label{fig:I_vs_beta}
\end{figure}

{To elucidate the role of spectral winding and motivate a suitable notion of classical response, we consider a generic} one-band non-Hermitian system with $N$ lattice sites, described by the following tight-binding Hamiltonian 
\begin{eqnarray}
H=\sum_{x=1}^N\sum_{j=-r}^{l} t_j\hat{c}^{\dagger}_j\hat{c}_{x+j},\label{eq:H_general}
\end{eqnarray}
with $t_j$ the hopping amplitude across $|j|$ lattice sites, $\hat{c}_x$ the annihilation operator of a (quasi-)particle at the $x$th lattice site, 
$\hat{c}_{x+N}=\hat{c}_x$ representing the PBC,
and $r$ ($l$) the maximal range of the hopping toward right (left) direction. 
The associated momentum-space Hamiltonian is given by
\begin{equation}
H(z)=\sum_{j=-r}^{l} t_j z^j={P_{r+l}(z)}/{z^r},
\end{equation}
with $k$ the quasi-momentum, $z:=e^{ik}$, and $P_{r+l}(z)$ a $(r+l)$th-order polynomial of $z$.
For any $t_j\neq t_{-j}^*$ (assuming $t_j=0$ if $j>l$ or $j<-r$), the Hamiltonian becomes non-Hermitian and possesses a point-gap topology, characterized by a nonzero spectral winding number w.r.t. a reference energy $E_r$ enclosed by the PBC spectrum~\cite{gong2018topological,okuma2020topological,zhang2019correspondence},
\begin{equation}
\nu(E_r)=\oint_\mathcal{C}\frac{dz}{2\pi }\frac{d}{dz}{\rm arg}\det[H(z)-E_r],\label{eq:winding}
\end{equation}
with $\mathcal{C}$ being the Brillouin zone (BZ), i.e. $k$ varying from $0$ to $2\pi$.
Simply put, $\nu(E_r)$ gives the number of times the PBC spectrum winds around $E_r$, as illustrated by a representative example in Fig.~\ref{fig:I_vs_beta}(a), corresponding to the Hamiltonian of Eq. \ref{eq:H_general} with $r=l=2$ .
As a side note, $\nu(E_r)$ also reflects the degeneracy of eigenmodes at $E_r$ when the system is placed under {semi-infinite boundary conditions} (SIBC) \GG{\cite{okuma2020topological}}.

Unlike the loop-like PBC spectrum depicting a nontrivial point-gap topology, the OBC spectrum must not cover any finite area in the complex plane, and so generically \GG{must take the form of curves within the PBC spectral loops} \cite{xiong2018does,Lee2019anatomy}, e.g. the Y-shape lines formed by the gray dots in Fig.~\ref{fig:I_vs_beta}(a).
That is, any reference energy $E_r$ inside a PBC spectral loop is enclosed $\nu(E_r)$ times by the PBC spectrum as $k$ varies from 0 to $2\pi$, but the same $E_r$ cannot be enclosed by the OBC spectrum. The important qualitative insight is hence the following: if we continuously deform the system from PBC to OBC, the evolving spectrum gradually collapses from the PBC loop spectrum to the OBC line spectrum, and is therefore expected to pass $E_r$ for $\nu(E_r)$ times.

To further appreciate this understanding, we consider the real-space Hamiltonian of Eq.~(\ref{eq:H_general}) with the following substitution only at the system's boundary, i.e., 
$t_j\rightarrow e^{-\beta} t_j$,
if $x+j>N$ or $x+j<1$.  This introduces an additional scaling factor, or an impurity,  to the boundary couplings of a 1D chain with $N$ unit cells. 
If one now continuously varies $\beta$ from $0$ to infinity, a PBC-OBC spectral evolution can be examined in detail. 
 
As shown in Fig.~\ref{fig:I_vs_beta}(a), by considering only $t_{\pm 1}$ and $t_{\pm 2}$,  we already have a  representative and intriguing model  whose spectral winding number can be either 1 or 2 in the topologically nontrivial regime. Next we consider the tuning of the boundary coupling via $t_{\pm1}\rightarrow e^{-\beta} t_{\pm1}$ for $x=N$ and $t_{\pm2}\rightarrow e^{-\beta} t_{\pm2}$ for $x=N,N-1$.
Let the $n$-th right eigenvalue  of this model system with such PBC-OBC interpolations be $E_n(\beta)$.  Then the spectral evolution is all captured by $E_n(\beta)$ vs $\beta$.  
{To motivate the connection with the Green's function, and to capture incidences when $E_n(\beta)$ comes close to $E_r$, we also define a quantity 
\begin{equation}
I_\beta(E_r)=\sum_n |1/(E_n(\beta)-E_r)|,
\label{Ibeta}
\end{equation}}
i.e. the absolute sum of 
the inverse energy spacings between the evolving spectrum $E_n(\beta)$  and a reference energy $E_r$.  
For an actual system always of finite size,  $E_n(\beta)$ is discretized, but it can still be made to be very close to the PBC reference eigenvalue $E_r$ for a sufficiently large system. 
As such, the quantity $I_\beta(E_r)$ can be a diagnosis tool to examine how many
times  $E_n(\beta)$  visits (the proximity of) $E_r$ as $\beta$ is tuned.  Moreover, the OBC limit can be essentially reached once $\beta$ is beyond a critical value $\beta=\beta_{\rm OBC}\sim {N\eta}$ with $\eta$ the effective localizing length of the eigenstates \cite{koch2020bulk,li2020impurity}.  With these understandings,  one infers that as $\beta$ varies from $0$ to $\beta_{\rm OBC}$, $I_\beta(E_r)$ is expected to display high peaks whenever the complex spectral evolution passes through $E_r$.  As explained above, the total number of such local peaks then reflects the spectral winding number $\nu(E_r)$. 
In Fig. \ref{fig:I_vs_beta}(b) we illustrate $I_\beta(E_r)$ as a function of $\beta$ for several $E_r$ 
denoted by the stars of different colors in Fig.~\ref{fig:I_vs_beta}(a), corresponding to different spectral winding numbers $\nu(E_r)$.  It is indeed  observed that the number of peaks of $I_\beta(E_r)$ directly reflects the spectral winding number $\nu(E_r)$. 
In the Methods section, we offer more insights based on the so-called generalized Brillouin zone (GBZ), to better understand
why $\nu(E_r)$ can be captured by the number of singularities encountered throughout the complex spectral evolution.


\subsection{Quantized response in signal amplification}

{While the previously defined quantity $I_\beta(E_r)$ is useful in diagnosing the spectral winding, it is not directly measurable. Below, we show how it inspires another analogous quantity that is directly associated with signal amplification.} 
\GG{In particular,  the quantity introduced below displays quantized plateaus that precisely match spectral winding numbers, making it possible to distinguish between one nontrivial point-gap topology from another.  This is a true advance as compared with earlier interesting attempts where signal amplification was only used to probe NHSE under OBC~\cite{Wanjura2020,xue2020non,yi2020nonH}. }
For our system with the PBC-OBC interpolation parameter $\beta$ ($0 <\beta<\beta_{\rm OBC}$),  what enters into the expression of the {Green's function $G$ in Eq. \ref{eq:green_sr}} is $1/[E_r-E_n(\beta)]$ again (as in $I_\beta(E_r)$)  and hence the Green's function should be able to capture the complex spectral evolution.  More importantly, it is found that the associated eigenmodes under PBC-OBC interpolation pile up at its boundary in a manner qualitatively different from that of NHSE  \cite{li2020impurity}. 
{With further derivation in the Method section, }
we find 
\begin{eqnarray}
|G_{1N}|\propto e^{\beta},~\frac{d\ln |G_{1N}|}{d\beta}=1
\label{G1N}
\end{eqnarray}
for a one-dimensional system with only the nearest-neighbor couplings, provided $t_1>t_{-1}$ and $E_r$ falls within the loop-spectrum of the system at a given $\beta$ (See Supplemental Materials).
For $t_{-1}>t_{1}$, one can analogously obtain $\frac{d\ln |G_{N1}|}{d\beta}=1$, corresponding to a winding number $\nu(E_r)=-1$. {We emphasize that the variation in Eq.~\ref{G1N} can be directly measured in a steady-state response experiment. In an electrical circuit setting, this task can be done via impedance measurements, as will be elaborated later.}

For more general cases with couplings across $r$ ($l$) lattice sites to the right (left), the system can be effectively understood as $m={\rm Max}[r,l]$ different sub-chains, with $t_{\pm m}$ viewed as the nearest neighbor coupling on each sub-chain, and the rest understood as inter-sub-chain or intra-chain longer-range couplings. For example, lattice sites $j$, $j+m$, $j+2m$, $\cdots$, with each pair of neighbors coupled by $t_{\pm m}$, form the $j$-th sub-chain, for $j=0, 1, 2, \cdots, m-1$. 
The above picture based on $m$ sub-chains becomes most useful if $t_{\pm m}$ is much larger than all other hopping parameters. Under this assumption  these sub-chains are nearly decoupled from each other in the bulk, and their respective edges are connected through the couplings modified by $e^{-\beta}$.  
This then suggests a unit-cell structure with $m$ sublattices, which greatly simplifies our consideration of signal amplification.  Indeed,   the overall amplification can be captured by the $m\times m$ block at the top-right (bottom-left) corners of the overall Green's function $G$, denoted as $G_{\leftarrow,m\times m}$ ($G_{\rightarrow,m\times m}$), corresponding to measuring the output at the first (last) $m$ sites of a signal entering from the last (first) $m$ sites. 
Since each sub-chain yields an amplification factor proportional to $e^{\beta}$, one qualitatively expects that $|G_{\leftarrow,m\times m}|\equiv \det[G_{\leftarrow,m\times m}]\propto e^{m\beta}$ or $|G_{\rightarrow,m\times m}|\equiv \det[G_{\rightarrow,m\times m}]\propto e^{m\beta}$, depending on the amplification direction.  This indicates that {
\begin{equation}
\nu_{\leftarrow,m}\equiv d\ln |G_{\leftarrow,m\times m}|/d\beta
\label{nuu}
\end{equation}
or analogously }$\nu_{\leftarrow,m}\equiv d\ln |G_{\leftarrow,m\times m}|/d\beta$ is quantized at $m$. 
On the other hand, the PBC spectrum of the system winds $m$ times around the origin of the complex plane, a fact obviously true since the associated momentum-space Hamiltonian is dominated by the terms $e^{\pm i mk}$. 
 As theoretically shown in Supplementary Material \cite{suppmat},  this correspondence between quantized response and spectral winding indeed holds,  if $E_r$ falls inside the PBC spectrum loop with the spectral winding number $\nu(E_r)$ given by $\pm m$, where the $\pm$ sign is to be determined by the actual direction of amplification.



\begin{figure}
\includegraphics[width=1\linewidth]{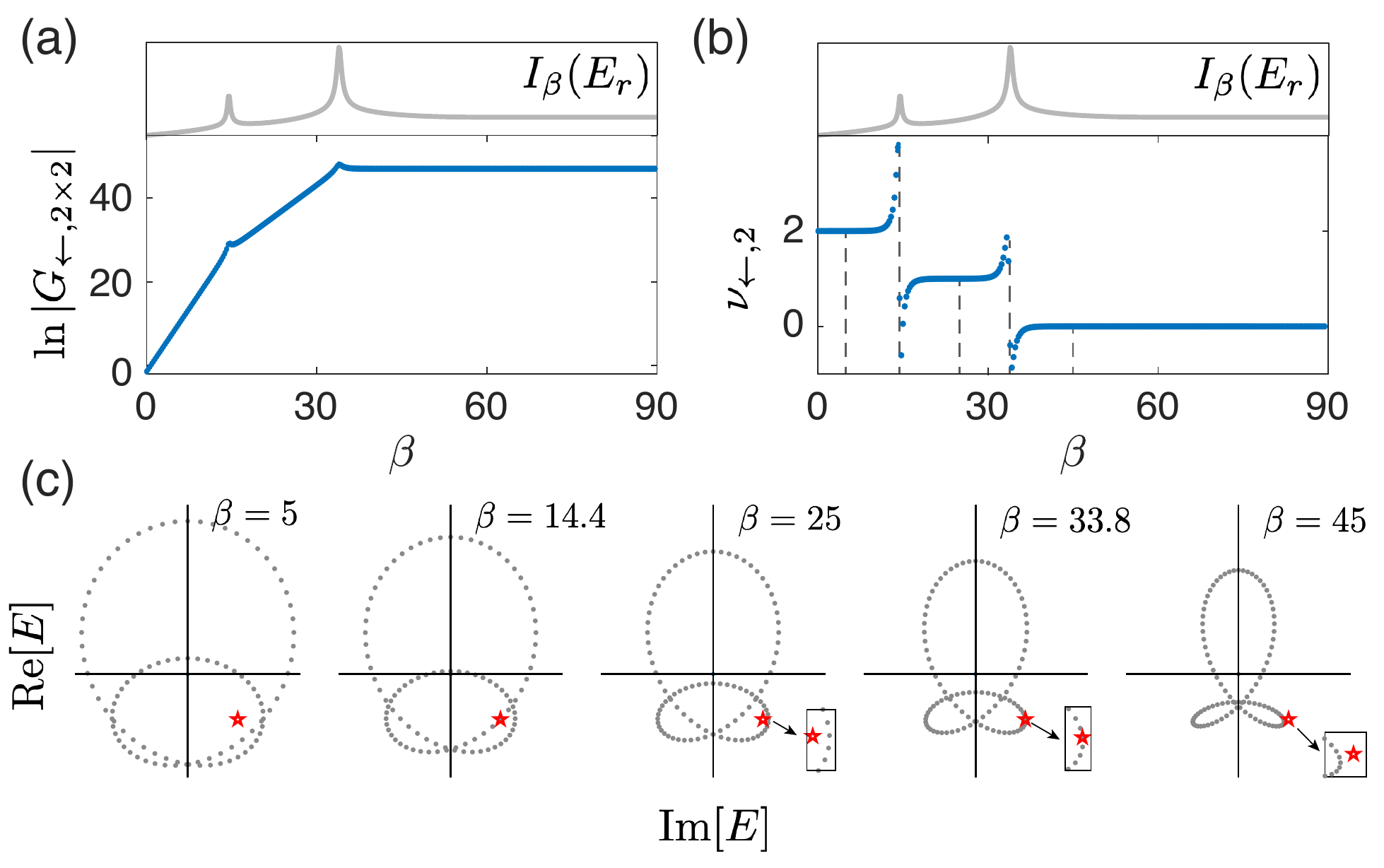}
\caption{
(a) Amplification ratio $|G_{\leftarrow,2\times 2}|$ and (b) $\nu_{\leftarrow,2}$,  the  derivative of $\ln|G_{\leftarrow,2\times 2}|$ over $\beta$, as functions of $\beta$, along with $I_\beta(E_r)$, whose peaks match well with the jumps between different plateaus.
(c) Spectrum at different values of $\beta$, corresponding to the five dashed lines in (b), respectively.
Red stars indicate the reference energy $E_r=\omega+i\gamma$, with $\omega=-0.96$ and $\gamma=1$.
Insets zoom in on the regime around $E_r$ to give a clearer view of the relation between the shown spectrum and $E_r$.
Parameters are set at $t_1=1$, $t_{-1}=0.5$, $t_2=2$, $t_{-2}=0$, and $N=100$.
}
\label{fig:Green}
\end{figure}

Figure \ref{fig:Green} presents computational results for $\ln \left|G_{\leftarrow,2\times 2}\right|$ and its derivative with respect to $\beta$ as functions of $\beta$, denoted $\nu_{\leftarrow, 2}$, again for the same system as that in Fig. \ref{fig:I_vs_beta}.  We also compare these results with $I(E_r)$ defined previously for a chosen reference energy point $E_r=\omega+i\gamma$, with $\nu(E_r)=2$.
It is observed that $\nu_{\leftarrow, 2}$ as a measurable physical response
shows three clear plateaus quantized at $\nu_{\leftarrow,2}=2,1,0$. Echoing with the jumps between these plateaus during the spectral evolution,  $I_{\beta}(E_r)$ shows local peaks whenever the spectral evolution passes through $E_r$.  
As shown in  Fig.~\ref{fig:Green}(c),
these transitions during the spectral evolution as a result of increasing $\beta$ match precisely with the $\beta$ values for which the complex spectrum touches the reference energy point and hence the spectral winding number is about to jump.  That is, the transitions between these quantized plateaus have a clear topological origin and are hence identified as topological transitions.  As a side remark,  observing all the plateaus in our example here happens to require a broad regime of $\beta$ and hence cases with very weak boundary coupling.  This is however not a concern because the first plateau at small $\beta$ is the one to reflect the topology under PBC.

\begin{figure}
\includegraphics[width=1\linewidth]{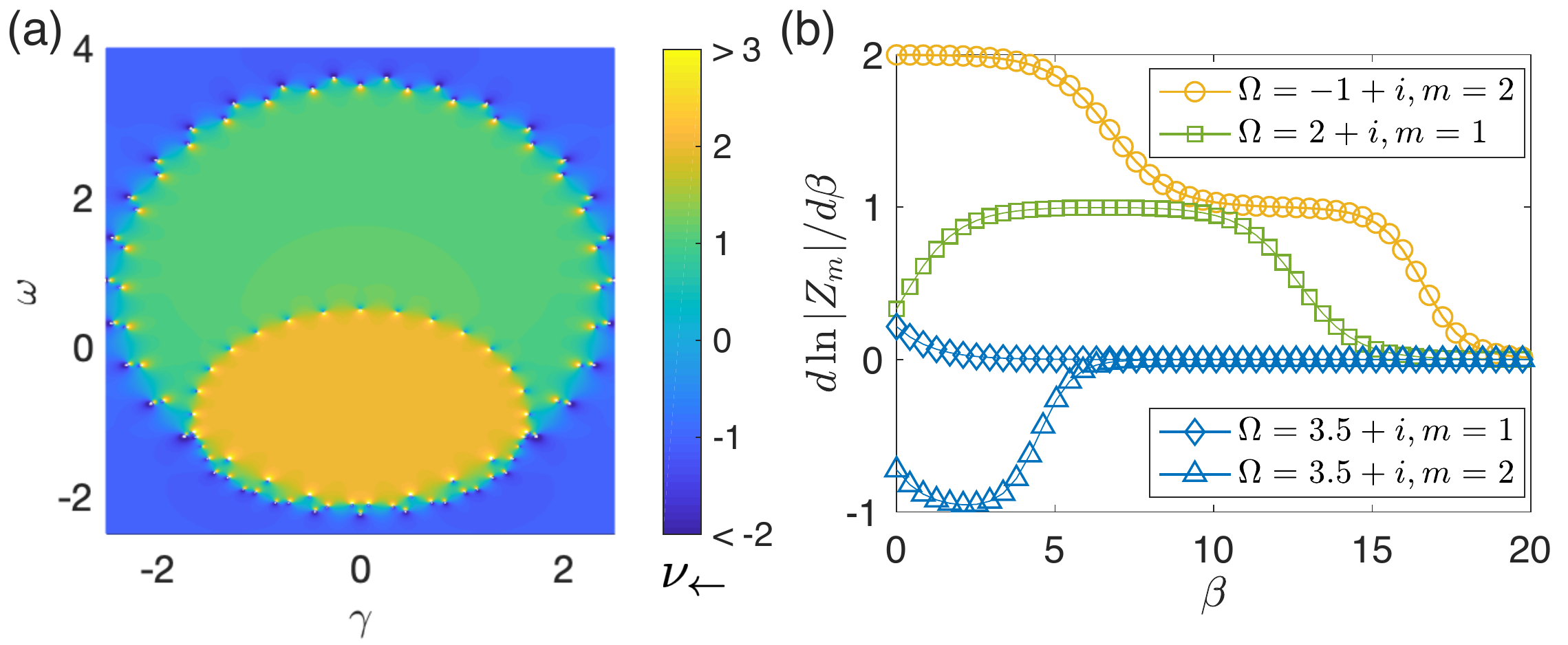}
\caption{The topological phase diagram and simulation of impedance measurement in a circuit realization.
(a) The topological phase diagram, in excellent agreement with that shown in Fig.~\ref{fig:I_vs_beta} is mapped out from examining physical response functions  $\nu_{\leftarrow,m}$ with $m=1$ or $m=2$, for the model depicted in the main text, with parameters set at  $t_1=1$, $t_{-1}=0.5$, $t_2=2$, $t_{-2}=0$, and $N=100$. 
 At the phase boundaries, $E_r=\omega+i\gamma$ is an eigenvalue of the system, leading to the divergence of the Green's function matrix.  
(b) Quantization of $\frac{d \ln| Z_{m}|}{d\beta}$, which is directly obtainable from simulated circuit measurement data of the impedance between the first $m$ and last $m$ nodes of the circuit lattice, here computed with $50$ nodes.
$Z_{m}$ (Eq.~\ref{Zm2} is the $m\times m$ matrix of impedances between the measured nodes, and $\beta$ is the parameter controlling the strengths of the couplings connecting the two ends.
\GG{The complex admittance parameter $\Omega$ indicated on panel (b) is connected with parameters $\omega$ and $\gamma$ plotted in panel (a) with $\omega={\rm Re}(\Omega)$ and $\gamma={\rm Im}(\Omega)$.}
 {The first plateau encountered when $\beta$ is increased from zero gives the {nonzero} topological winding $\nu$ (green, yellow for $\nu=1,2$). } 
{When $\nu=0$, the logarithmic gradients of both $Z_{1,2}$ are close to or smaller than $0$ (blue), corresponding to the absence of signal amplification toward either direction.} 
}
\label{fig:G_diagram}
\end{figure}

The results presented in Fig.~\ref{fig:Green} are particularly stimulating. Indeed, therein neither the  next-nearest-neighbor coupling nor the nearest-neighbor coupling is dominating.  Yet quantized plateaus at $m=1$ and $m=2$ are still obtained.   Returning to our decoupled sub-chain picture above, this indicates that for different reference energy points, the behavior of this system is topologically equivalent to that of one single chain or that of two weakly coupled sub-chains.  With this perspective, we propose to examine $\nu_{\leftarrow,m}$ vs. different choices of $m$ in order to fully map out the phase boundaries, without any prior knowledge of spectral winding.    We thus proceed to find the maximal $\nu_{\leftarrow,m}$ by scanning $m$, for sufficiently small $\beta$.  The obtained value is then expected to yield $\nu(E_r)$ of the studied system under PBC.   To this end we define $\nu_\leftarrow={\rm Max}[\nu_{\leftarrow,1},\nu_{\leftarrow,2}]$, given that our model system at most has effectively two sub-chains (one can similarly define $\nu_\rightarrow$).
We present in Fig.~\ref{fig:G_diagram}(a) our results of $\nu_\leftarrow$ at $\beta=0$ for different $\omega$ and $\gamma$ , with the phase boundaries identified there in excellent agreement  with the actual PBC spectrum shown in Fig.~\ref{fig:I_vs_beta}(a). The only subtlety is that we also obtain a negative value
 $\nu_\leftarrow=-1$ in the topological trivial regime (theoretically this result is also explained in Supplementary Material \cite{suppmat}).   However, in these regimes  both $\nu_\leftarrow$ and $\nu_\rightarrow$ are found to be negative, the amplification factor is far less than unity, and hence there is actually no signal amplification after all.  One may just exploit this additional feature to locate regimes with zero spectral winding.

\subsection{Measurement of quantized response in electrical circuits}

We next elaborate on how the quantized response can be directly extracted by measuring the impedance in an electrical circuit setting. Instead of external perturbations, a circuit is most naturally driven by a steady-state AC or DC current, with voltage response given by Kirchhoff's law  $\bold I = L\bold V$, where $L$ is the circuit Laplacian matrix and the components of $\bold I$ and $\bold V$ are respectively the input currents and voltages at each node. 
Suppose that the circuit is then grounded by identical circuit components with complex admittances $-\Omega$. In this case, the full (grounded) Laplacian becomes $J=L-\Omega \,\mathbb{I}$, and the voltage distribution due to the input current are given by~\cite{lee2018topolectrical,lee2020imaging}
\begin{equation}
\bold V_i = [J^{-1}]_{ij}\bold I_j =  G_{ij}\bold I_j=\left[\sum_n\frac{|\Psi^R_n\rangle\langle \Psi_n^L|}{\Omega-E_n}\right]_{ij}\bold I_j
\end{equation}
where the eigenvalues $E_n$ and L/R eigenstates $|\Psi^{L/R}_n\rangle$ are that of the Laplacian $J$. Notably, the quantity in the square parentheses agree exactly with our definition of the Green's function (Eq.~\ref{G}), with $\Omega$ taking the role of $\omega+i\gamma$. We extract the physical response through the impedance $Z_{ij}$ between the $i$-th and $j$-th nodes, which is related to the Green's function via $Z_{ij}=G_{ii}+G_{jj}-G_{ij}-G_{ji}$. By varying the identical grounding admittances $\Omega$ via a combination of RLC components with $\pm\pi/2$ relative phase shifts, we will be able to effectively access the response from different regions of the complex spectral plane. 

Quantized classical response can be obtained from a circuit whose Laplacian exhibits nontrivial spectral winding. This requires effectively asymmetric couplings between nodes, which has already been demonstrated in existing topolectrical experiments through combinations of capacitors, inductors and INICs (negative impedance converter with current inversion) comprising operation amplifiers~\cite{hofmann2019chiral,helbig2020generalized,ezawa2019electric}, as further elaborated in the Supplementary Material. The boundary couplings can also be adjusted to effect the variation with $\beta$ through tunable inductors connected in series with the asymmetric couplings~\cite{li2020impurity}. Arbitrarily large spectral winding numbers can always be achieved by coupling sufficiently distant nodes, which can be much more feasibly done in electrical circuits compared to other platforms.

In analogy to $\nu_{\leftarrow,m}$ from Eq.~\ref{nuu}, we can define $Z_m=\text{det}\,Z_{ij}|_{i\leq m, N-j<m}$, the determinant of the $m\times m$ matrix of impedances between the first $m$ nodes at one end with the last $m$ at the other end of the circuit chain. Although it is not exactly equivalent to $\nu_{\leftarrow,m}$, 
it is expected to vary with $\beta$ in a similar manner, since the impedance $Z_{ij}$ is dominated by the component of the Green's function that produces the directional amplification. Therefore,
by keeping track of the effective $\beta(\omega)$ and $\Omega(\omega)$, the logarithmic gradient of the impedance determinant $\frac{d \ln| Z_{m}|}{d\beta}$ will be expected to exhibit quantized jumps as $\Omega(\omega)$ crosses the boundary between regions of different topological winding $\nu$.  For $m\leqslant 2$, which includes the model we had considered (Figs.~1, 2 and 3), we explicitly have 
\begin{equation}
Z_{m=2}=Z_{1,N-1}Z_{2,N}-Z_{1,N}Z_{2,N-1},
\label{Zm2}
\end{equation}
whose gradients are dominated by those of terms like $G_{1,N-1}G_{2,N}$ and $G_{1,N}G_{2,N-1}$ in the presence of directional amplification (toward the first lattice site). As demonstrated via the simulated measurements in Fig.~\ref{fig:G_diagram}(b), the gradient $\frac{d \ln| Z_{m=1,2}|}{d\beta}$ indeed exhibits plateaus quantized at the winding number $\nu$ (blue, green, yellow for $\nu=0,1,2$ respectively) where $\Omega$ is tuned to. For higher topological winding, we take the plateau that first occurs when $\beta$ is increased from $0$, i.e. the plateau closest to periodic boundary conditions. 

\section{Conclusions}
{In this work, we have introduced the new paradigm of quantized classical response, where a quantized response coefficient can always be extracted from how the Green's function \GG{varies} with the evolution of an imaginary flux-like parameter.
 Being based on the topological winding properties of the Green's function in the complex energy plane, \GG{this quantization} does not assume the existence of any quantum mechanical ground state, and applies to all systems, classical and quantum.
Specifically, we show that the spectral winding number is directly detectable as a steady-state response coefficient to changes in the boundary condition.  \GG{Such correspondence between spectral winding numbers and quantized response is arguably broader in scope than in the case of momentum-space topology, because spectral winding does not even require translational invariance.} \GG{Our results are relevant to a number of current experimental platforms of non-Hermitian systems \cite{hofmann2020reciprocal, helbig2020generalized,xiao2020non,Wanjura2020,li2020critical,NPReview}.}
 In the context of classical electrical circuits, we have shown that a quantized response can be easily extracted from extremely experimentally accessible impedance measurements.


}

\begin{acknowledgements}
{\it Acknowledgement.-}
We would like to thank Da-Jian Zhang for helpful discussions.  
L. L. acknowledges funding support by
the Key-Area Research and Development Program of GuangDong Province under Grants No. 2019B030330001 
and the Startup Grant of Sun Yat-sen University (No. 71000-18841245).
J.G. acknowledges funding support by the Singapore NRF
Grant No. NRF-NRFI2017-04 (WBS No. R-144-000-378-281). C.H.L acknowledges funding support by the Singapore MOE Tier-1 start-up grant (WBS R-144-000-435-133).
\end{acknowledgements}

\section{Methods}
\subsection{Insights based on generalized Brillouin zone}
\begin{figure}
\includegraphics[width=0.7\linewidth]{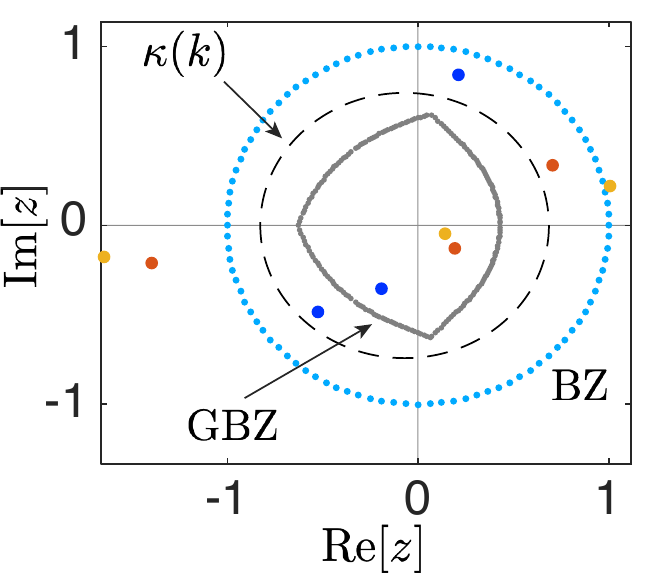}
\caption{A typical example of the Brillouin zone (BZ) and the generalized Brillouin zone (GBZ).
The BZ and GBZ are given by cyan and gray dots respectively, in the complex plane of $z:=e^{ik}e^{-\kappa(k)}$. Here $\kappa(k)$ represents a complex deformation of the momentum $k$.
The black dash loop indicates an evolving GBZ with $\kappa(k)$ between $0$ and $\kappa_{\rm OBC}(k)$, the value that gives the spectrum under the open boundary condition. Blue, red, and yellow dots are the zeros of $H(z)-E_r$, where $H(z)$ is the PBC Hamiltonian of the system and $E_r$ is the chosen reference energy for calculating the winding number. The system is chosen as the same as that in Fig. \ref{fig:I_vs_beta} in the main text,
i.e. $H(z)=2z^2+z+1/2z$.
}
\label{fig:GBZ}
\end{figure}

\GG{Here} we offer more insights based on the so-called generalized Brillouin zone (GBZ), to better understand
why $\nu(E_r)$ can be captured by the complex spectral evolution.
According to the non-Bloch band theory, the OBC spectrum can be described by the PBC one in a GBZ, using a complex deformation of the quasi-momentum $k\rightarrow k+i\kappa_{\rm OBC}(k)$ \cite{yao2018edge,Lee2019anatomy,lee2018tidal,lee2020unraveling,yokomizo2019non}.
The PBC-OBC spectral evolution can then be effectively described by $k\rightarrow k+i\kappa(k)$ with $\kappa(k)$ varying from $0$ to $\kappa_{\rm OBC}(k)$, with $\kappa_{\rm OBC}(k)$ having the minimal magnitude to yield  the OBC spectrum \cite{yokomizo2019non,Lee2019anatomy}.   The PBC-OBC spectral evolution can hence be understood as arising from tuning $\kappa(k)$ and hence deforming the BZ to the GBZ, as shown in Fig.~\ref{fig:GBZ}. 
Moreover, an winding number can be defined as,
\begin{equation}
\nu(E_r)=\oint_{\rm GBZ}\frac{dz}{2\pi }\frac{d}{dz}{\rm arg}\det[H(z)-E_r],\label{eq:winding_GBZ}
\end{equation}
analogous to that of Eq.~(\ref{eq:winding}), with integration in the GBZ instead of the BZ.
This winding number must be zero when the OBC spectrum is reached because, again, the OBC spectrum cannot enclose any finite area \cite{okuma2020topological,zhang2019correspondence}.  
Following the Cauchy principle,  the spectral winding number is found to be
$\nu(E_r)=N_{\rm zero}-N_{\rm pole}$,
where $N_{\rm zero}$ and $N_{\rm pole}$ are the counting of zeros and poles enclosed by the integration path (BZ or GBZ) weighted by their respective orders.
The conclusion is hence as simple as follows. If we continuously tune $\kappa(k)$, the PBC-OBC spectral evolution must pass through different zeros of $[\frac{P_{r+l}(z)}{z^r}-E_r]$ [colored dots in Fig.~\ref{fig:GBZ}] for a total of  $\nu(E_r)$ times, such that the spectral winding number reduces from $\nu(E_r)$ to $0$ eventually when the integration path approaches the GBZ.
Thus, during the complex spectral evolution,  the spectrum under $k\rightarrow k+i\kappa(k)$ must pass the reference energy $E_r$ for a total of $\nu(E_r)$ times, constituting a rather formal argument to justify our treatment in the main text.

\subsection{Directional signal amplification versus the PBC-OBC spectral evolution}
\subsubsection{Cases with only nearest-neighbor coupling}
As mentioned in the main text, under an external drive $\vec{\epsilon}(t)=\vec{\epsilon}(\omega){\rm exp}(-i\omega t)$ and an overall on-site gain/loss parameter $\gamma$, the resultant response field $\vec{\phi}(t)$ can be written as $\vec{\phi}(t)=\vec{\phi}(\omega){\rm exp}(-i\omega t)$, with
\begin{eqnarray}
\vec{\phi}(\omega)=G(\omega,\gamma)\vec{\epsilon}(\omega),~G(\omega,\gamma)=\frac{1}{E_r-H},
\end{eqnarray}
where $E_r=\omega+i\gamma$, $G$ is the Green's function matrix. 
The amplification factors for a signal toward the left and the right are described by the matrix elements $G_{1N}$ and $G_{N1}$, respectively.
For a non-Hermitian $H$, the Green's function matrix can be expressed in the spectral representation \cite{zirnstein2019bulk,xue2020non}
\begin{eqnarray}
G(\omega,\gamma)=\frac{1}{E_r-H}=\sum_n\frac{1}{E_r-E_n^R}|\Psi^R_n\rangle\langle\Psi^L_n|,
\end{eqnarray}
with $|\Psi^R_n\rangle$ the $n$th right eigenstate of $H$ with eigenenergy $E_n^R$, and $\langle\Psi^L_n|$ the corresponding left eigenstate.

To be more explicit, consider the Hatano-Nelson model under the PBC-OBC interpolation, described by the following Hamiltonian
\begin{equation}
  H_\beta = \sum_{x=1}^{N-1}(t_1\hat{c}^{\dagger}_{x}\hat{c}_{x+1}+t_{-1}\hat{c}^{\dagger}_{x+1}\hat{c}_x)
  +e^{-\beta}(t_{1}\hat{c}^{\dagger}_{N}\hat{c}_{1}+t_{-1}\hat{c}^{\dagger}_{1}\hat{c}_N),
  \label{hn_model}
\end{equation}
also with the assumption $t_1>t_{-1}$ without loss of generality.
Let the $n$-th right eigenstate be $|\Psi_n^R\rangle=\sum_{x=1}^{\infty}\psi_{x,n}^R\hat{c}^{\dagger}_{x}|0\rangle$, with $|0\rangle$ the vacuum state.  Using the eigenvalue-eigenstate equation $H_\beta |\Psi_n^R\rangle=E_n^R|\Psi_n^R\rangle$, one obtains
\begin{eqnarray}
t_{1} \psi^R_{x+1,n}+t_{-1} \psi^R_{x-1,n}=E_n^R\psi^R_{x,n}~\label{eq:bulk_HN} 
\end{eqnarray} for $x\in[2,N-1]$,
and
\begin{eqnarray}
e^{-\beta}t_{1} \psi^R_{1,n}+t_{-1} \psi^R_{N-1,n}&=&E_n^R\psi^R_{N,n},\label{eq:boundary1_HN}\\
t_{1} \psi^R_{2,n}+e^{-\beta}t_{-1} \psi^R_{N,n}&=&E_n^R\psi^R_{1,n}.\label{eq:boundary2_HN}
\end{eqnarray}
If we further assume $t_1\gg t_{-1}$ and the following exponentially decaying eigensolutions:
\begin{eqnarray}
\psi_{x,n}^R=C_n e^{-M_n^R (x-1)}
\end{eqnarray}
with $C_n$ the normalization constant and $M_n^R>0$,  then one obtains
\begin{eqnarray}
t_{1} e^{-M_n^R}=E_n^R, M_n^R=\frac{\beta-i2n\pi}{N}.
\label{loopR}
\end{eqnarray}
Note that here $|E_n^{R}/t_1|$ is determined by the ratio $\beta/N$.  That is, the eigenvalues $E_n^{R}$ will be distributed on a circle on the complex plane, whose radius depends only on $t_1$ as well as the ratio  $\beta/N$.

Likewise,  the left eigenstates under the same assumptions satisfy
$H^\dagger_\beta|\Psi_{n}^L\rangle=E_n^L|\Psi_{n}^L\rangle$ and $E_n^L=(E_n^R)^*$,
and they are found to be
\begin{eqnarray}
\psi_{x,n}^L&=&C_n^* e^{-M_n^L (N-x)}, t_1 e^{-M_n^L} =E_n^L, \nonumber\\
M_n^L&=&\frac{\beta+i2n\pi}{N}.
\end{eqnarray}
From the biothorgonal condition $\langle \Psi_{n}^L|\Psi_{n}^R\rangle=1$, we then obtain the normalization constant
\begin{eqnarray}
C_n=\frac{e^{-i\frac{N-1}{N}\pi n}}{\sqrt{Ne^{-\beta(N-1)/N}}}.
\end{eqnarray}

\begin{figure*}
\includegraphics[width=1\linewidth]{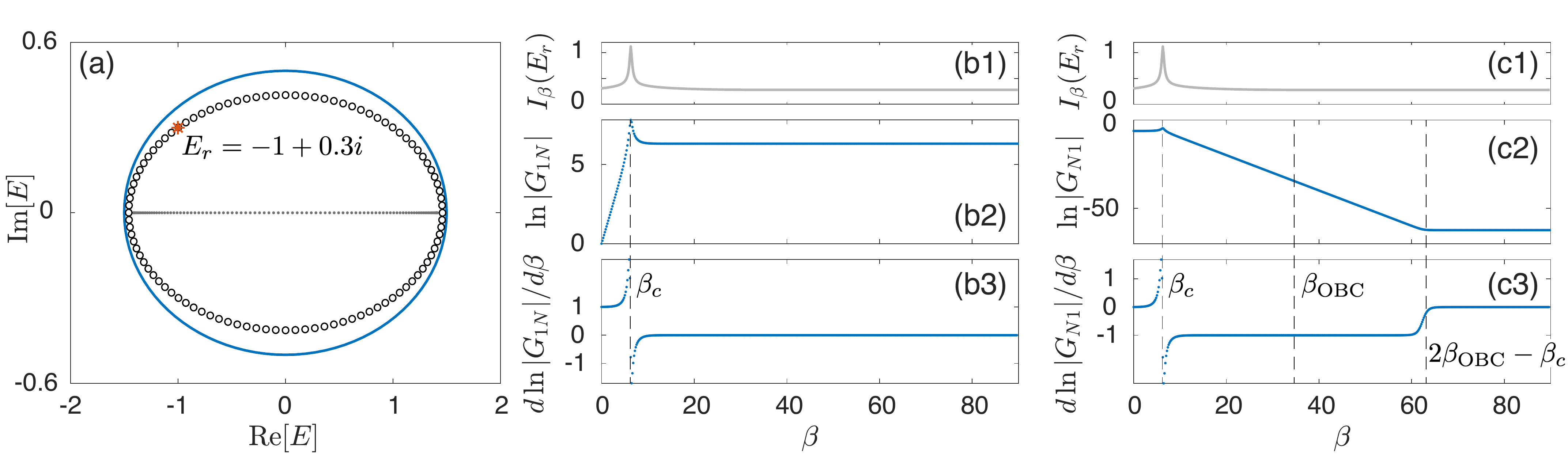}
\caption{(a) Spectra under PBC (cyan loop), OBC (gray dots), and (black circles) the PBC-OBC interpolation through $\beta$. The red star indicates the reference energy $E_r=\omega+i\gamma$ for further results in (b) and (c). Here $E_r$ falls right on the spectrum of $H_\beta$, as the value of $\beta=\beta_c\approx 6.2$ is read out from (b) and (c).
(b) and (c) show the amplification ratios $|G_{1N}|$ and $|G_{N1}|$ [(b2) and (c2)] and the derivatives of their logarithms over $\beta$ [(b3) and (c3)] as functions of $\beta$, where the quantity $I_\beta(E_r)$ indicates with its peaks when $E_r$ is passed through by the $H_\beta$ spectrum.
The parameters are $t_1=1$ and $t_{-1}=0.5$.
}
\label{fig:G_NN}
\end{figure*} 

With preparations above the matrix element of the Green's function $G$ can then be found as follows:
\begin{eqnarray}
G_{1N}&=&\sum_n\frac{1}{E_r-E_n^R}\psi_{1,n}^R(\psi_{N,n}^L)^*\nonumber\\
&=&\sum_n\frac{1}{E_r-t_1 e^{-\frac{\beta-i2n\pi}{N}}}
\frac{1}{N (e^{-\frac{\beta-i2n\pi}{N}})^{N-1}}\label{eq:G1N_1}\\
&=&\sum_n\frac{e^{\beta}e^{-\frac{\beta-i2n\pi}{N}}}{E_r-t_1 e^{-\frac{\beta-i2n\pi}{N}}}
\frac{1}{N}\label{eq:G1N_2}.
\end{eqnarray}
We may now attempt to rewrite the discrete sum in Eq.~(\ref{eq:G1N_2}) in terms of a loop integral of a complex variable $z$, by defining $k_n:=2n\pi/N$ and $z:=e^{-\beta/N}e^{ik_n}$. 
To that end one must implicitly assume that $N$ under consideration is sufficiently large so as to use an integral to replace the discrete sum.  With this in mind, for a given $\beta$, $\beta/N$ is assumed to be vanishingly small, and hence essentially we are working in the regime of $|z|\rightarrow 1$.  Under these conditions, the sum in Eq.~(\ref{eq:G1N_2}) can then be evaluated by the following integral
\begin{eqnarray}
G_{1N}=\oint_{|z|= e^{-\beta/N}} \frac{1}{2\pi i } \frac{e^\beta}{E_r-t_{1} z}dz,
\end{eqnarray}
which is found to be
\begin{eqnarray}
G_{1N} =- \frac{e^{\beta}}{t_1 }, \label{eq:G1N}
\end{eqnarray}
if $z_0 \equiv E_r/t_1$ satisfies $|z_0| <e^{-\beta/N}$, i.e. the pole of the integrand falls within the the integral loop. 
This condition leads to
\begin{eqnarray}
|E_r|<t_{1} e^{-\beta/N}=|E^R_n|,\label{eq:Er_enclosed}
\end{eqnarray}
meaning that the reference energy $E_r$ falls within the loop-spectrum of $H_\beta$.
The above detailed theoretical considerations indicate that,  so long as $E_r$ is enclosed by the loop spectrum of $H_\beta$, we have
\begin{eqnarray}
\frac{d \ln |G_{1N}|}{d \beta}=1,
\end{eqnarray}
which is just the claim in the main text regarding how to use a quantized physical response to detect the spectral winding number $\nu(E_r)$, as computationally verified in Fig.~\ref{fig:G_NN}(b).

Next we investigate what happens if we let $\beta$ exceed $\beta_c$,  the critical $\beta$ value for which a reference energy point $E_r$ falls exactly on the loop spectrum of $H_\beta$ (as shown in Fig.~\ref{fig:G_NN}(a)).   Let us first recall the result from Eq.~(\ref{loopR}), which indicates that the radius of the loop spectrum of $H_\beta$ scales with $\beta/N$.   Specifically, for a given lattice size $N$ and a given reference point $E_r$ under investigation, one immediately obtains that 
\begin{eqnarray} 
\beta_c =- N \ln  |E_r/t_1|,
\label{betac}
\end{eqnarray} which is clearly proportional to $N$.  As such,  to probe the regime of  $\beta\geq \beta_c$ , $\beta$ should at least linearly increase with $N$ as well.   This being the case,  we can no longer approximate the discrete sum in Eq.~(\ref{eq:G1N_2}) as a loop integral with $N\rightarrow \infty$ in mind, simply because it has the factor $e^{\beta}$ in its numerator, which diverges with $N$ because in the regime of interest $\beta$ diverges with an increasing $N$. 

Based on the discussions above, in the regime of $\beta\geq \beta_c$, what is under investigation becomes the evaluation of the same discrete sum, but with $\beta$ scaling proportionally with $N$, and hence both becoming sufficiently large in the event of using a loop integral to replace this discrete sum.  Though we still use the same complex variable $z=e^{-\beta/N}e^{ik_n}$ to invoke a possible loop integral, we see that $|z|$ is now far from unity.  To reflect this, we now use the alternative expression of the discrete sum in Eq.~(\ref{eq:G1N_1}) and then rewriting it as the following integral, with the integrand having {one pole of order $N$ at $z=0$, and another $1$st-order pole at $z=E_r/t_1$, }
\begin{eqnarray}
G_{1N}&=&\oint_{|z|=e^{-\beta/N}} \frac{1}{2\pi i } \frac{1}{z^N(E_r-t_{1} z)}dz.
\label{exp2}
\end{eqnarray}
Assuming that $|z_0|=|E_r/t_1|>e^{-\beta/N}$,  namely, the reference energy $E_r$ is outside the loop, so $z=0$ is the only pole of the integrand {enclosed by the integration path}, we have
\begin{eqnarray} 
G_{1N}= \frac{t_1^{N-1}}{E_r^N}, \label{eq:G1N_3}
\end{eqnarray}
a value independent of $\beta$, which is again consistent with the computational results in Fig.~\ref{fig:G_NN}(b2).
Further, using the previous expression of $\beta_c$ from Eq.~(\ref{betac}), we find that in this case $G_{1N}=\frac{t_1^{N-1}}{E_r^N}=e^{\beta_c}/t_1$.   Interestingly, though this magnitude  $e^{\beta_c}/t_1$ exponentially larger than $e^{\beta}/t_1$ obtained earlier for $\beta<\beta_c$,  this amplification factor is saturated and no longer depends on $\beta$ in the regime of $\beta>\beta_c$.

As a side note, one might wonder why we cannot also use the loop integral in Eq.~(\ref{exp2}) to treat the first case, namely, 
a fixed $\beta$ in the regime of $\beta<\beta_c$ but with $N$ approaching sufficiently large values.   As said earlier, in this case we essentially perform the summation under the condition of $|z|=1$. Under this condition we always have $z^N=1$ and hence the expression in Eq.~(\ref{exp2}) is no longer useful.

In the same fashion, we can now proceed to examine $G_{N1}$, which depicts how the signal is amplified or suppressed  in the other direction. The matrix element $G_{N1}$ is found to be the following,
\begin{eqnarray}
G_{N1}&=&\sum_n\frac{1}{E_r-E_n^R}\psi_{N,n}^R(\psi_{1,n}^L)^*\nonumber\\
&=&\sum_n\frac{1}{E_r-t_1 e^{-\frac{\beta-i2n\pi}{N}}}
\frac{(e^{-\frac{\beta-i2n\pi}{N}})^{N-1}}{N }
\label{eq:GN1_1}\\
&=&\sum_n\frac{e^{-\beta}e^{\frac{\beta-i2n\pi}{N}}}{E_r-t_1 e^{-\frac{\beta-i2n\pi}{N}}}
\frac{1}{N}.\label{eq:GN1_2}
\end{eqnarray}
Rewriting Eq.~(\ref{eq:GN1_2}) in terms of a loop integral, we have
\begin{eqnarray}
G_{N1}&=&\oint_{|z|=e^{-\beta/N}} \frac{1}{2\pi i } \frac{e^{-\beta}}{z^2(E_r-t_{1} z)}dz.
\label{loop2}
\end{eqnarray}
The integrand has a 1st-order pole at $z_0=E_r/t_1$, and a second-order pole at $z_1=0$. Therefore we have
\begin{eqnarray}
G_{N1}= \frac{e^{-\beta}t_1}{E_r^2}
\end{eqnarray}
if $|E_r|>t_1e^{-\beta/N}$ (reference energy falls outside the loop of integral), and 
\begin{eqnarray}
G_{N1}= 0
\end{eqnarray}
if $|E_r|<t_1e^{-\beta/N}$ (reference energy $E_r$ falls inside the loop of integral). 
Here because of the factor $ e^{-\beta}$ in Eqs.~(\ref{eq:GN1_2}) and Eq.~({\ref{loop2}), the replacement of the discrete sum by the loop integral is always valid by assuming a sufficiently large $N$, i.e., regardless of whether $\beta$ is assumed to be fixed or assumed to scale linearly with $N$.  Thus,  results obtained above for both $\beta>\beta_c$ and $\beta<\beta_c$ are valid,  which are indeed consistent with our numerical results. 
{Note also that for fixed $\beta$, our numerical results for finite systems give a small but nonzero $G_{N1}$ when $\beta<\beta_c$ (e.g. $|G_{N1}|\approx e^{-5}$ in Fig.~\ref{fig:G_NN}(c2)), which vanishes when further increasing $N$ (not shown).}

Overall, we obtain that 
the gradient of $\ln|G_{N1}|$ with respect to $\beta$ is again quantized, with
\begin{eqnarray}
\frac{d \ln |G_{N1}|}{d \beta}=-1,
\end{eqnarray}
if $E_r$ is NOT enclosed by the loop spectrum of $H_\beta$,  corresponding to $\beta>\beta_c$ [Fig. \ref{fig:G_NN}(c)].
If we further increase $\beta$ the system shall approach the OBC limit when $\beta= \beta_{\rm OBC}\approx\alpha N$ with $\alpha=\ln (\sqrt{t_1/t_{-1}})$, {where the spectrum falls on the same lines as the OBC spectrum~\cite{kunst2018biorthogonal,koch2020bulk}. Nevertheless, this limit is still not exactly like the OBCs, as the two boundaries are still weakly connected. For example, due to the (weak) boundary couplings, a flux threading cannot be gauged away and can lead to fluctuation of eigenergies, unlike in real OBC cases.
Indeed, from our numerical results, we indeed see that $G_{N1}$ keeps decreasing after $\beta$ exceeds $\beta_{\rm OBC}$, and becomes a constant when $\beta\gtrsim 2\beta_{\rm OBC} -\beta_c$ [Fig. \ref{fig:G_NN}(c3)].

\subsubsection{Cases with only $m$th-nearest neighbor coupling}
Consider now a 1D non-Hermitian chain with only the $m$th-nearest neighbor couplings:
\begin{eqnarray}
  H_\beta &=& \sum_{x=1}^{N-m}(t_{m}\hat{c}^{\dagger}_{x}\hat{c}_{x+m}+t_{-m}\hat{c}^{\dagger}_{x+m}\hat{c}_x)\nonumber\\
  &&+e^{-\beta}\sum_{x=N-m+1}^{N}(t_{m}\hat{c}^{\dagger}_{x}\hat{c}_{x+m-N}+t_{-m}\hat{c}^{\dagger}_{x+m-N}\hat{c}_x).\nonumber\\
  \label{m_hn_model}
\end{eqnarray}
Here we first assume $N/m$ is an integer, 
thus the system is decoupled into $m$ identical 1D sub-chains, and the eigenstates satisfy
\begin{eqnarray}
t_{m} \psi^R_{x+m,n}+t_{-m} \psi^R_{x-m,n}&=&E_n^R\psi^R_{x,n}\label{eq:bulk_m_HN}
\end{eqnarray}
for $x\in[2,N-m]$, and
\begin{eqnarray}
e^{-\beta}t_{m} \psi^R_{s,n}+t_{-m} \psi^R_{N-2m+s,n}&=&E_n^R\psi^R_{N-m+s,n},\label{eq:boundary1_m_HN}\\
t_{m} \psi^R_{m+s,n}+e^{-\beta}t_{-m} \psi^R_{N-m+s,n}&=&E_n^R\psi^R_{s,n},\label{eq:boundary2_m_HN}
\end{eqnarray}
with $s=1,2,...,m$ labelling different sub-chains. 
As in the previous discussion for the case of $m=1$, we assume $t_m\gg t_{-m}$ to obtain some simple analytical results.
Here we replace the labels $x$ and $n$ with $x_s$ and $n_s$ for each sub-chain.  We take the ansatz
\begin{eqnarray}
\psi^R_{x_s,n_s}=C_{n_s}e^{-M_{n_s}^R(x_s-1)},
\end{eqnarray}
with $n_s$ only takeing values from $1$ to $N_m=N/m$, given that each sub-chain contains only $N_m$ lattice sites, and $x_s= (x-s+m)/m$ ranging from $1$ to $N_m$, with $x$ being $s,m+s,2m+s,...,N-m+s$, and more importantly, 
\begin{eqnarray}
t_{m} e^{-M_{n_s}^R}=E_{n_s}^R, M_{n_s}^R=\frac{\beta-i2n_s\pi}{N_m}.
\end{eqnarray}
Similarly, the left eigenstates are given by
\begin{eqnarray}
\psi_{x_s,n_s}^L&=&C_{n_s}^* e^{-M_{n_s}^L (N_m-x_s)}, t_{m} e^{-M_{n_s}^L} =E_{n_s}^L, \nonumber\\
M_{n_s}^L&=&\frac{\beta+i2{n_s}\pi}{N_m}.
\end{eqnarray}
Again, from the biothorgonal condition $\langle \Psi_{n}^L|\Psi_{n}^R\rangle=1$, we have the normalization constants
\begin{eqnarray}
C_{n_s}=\frac{e^{-i\frac{N_m-1}{N_m}\pi {n_s}}}{\sqrt{N_me^{-\beta(N_m-1)/N_m}}}.
\end{eqnarray}
Note that in the Green's function matrix, the element $G_{1N}$ shall always be zero as each sub-chain is decoupled from the others. 
In this case, the directional amplification of each sub-chain corresponds to the element
$G_{s(N-m+s)}$, with
\begin{eqnarray}
G_{s(N-m+s)}&=&\sum_{n_s}\frac{1}{E_r-E_{n_s}^R}\psi_{x_s=1,{n_s}}^R(\psi_{x_s=N_m,{n_s}}^L)^*\nonumber\\
&=&\sum_{n_s}\frac{1}{E_r-t_{m} e^{-\frac{\beta-i2n_s\pi}{N_m}}}
\frac{1}{N_m(e^{-\frac{\beta-i2n_s\pi}{N_m}})^{N_m-1}}\nonumber\\
\\
&=&\sum_{n_s}\frac{e^{\beta}e^{-\frac{\beta-i2n_s\pi}{N_m}}}{E_r-t_{m} e^{-\frac{\beta-i2n_s\pi}{N_m}}}\frac{1}{N_m}.
\end{eqnarray}
As $n_s$ takes value from $1$ to $N_m=N/m$, here we need to define $k_s=i2mn_s\pi/N$, so that the summation can be replaced by an integral with $k_s$ varying from $0$ to $2\pi$. Similar to the case with only nearest-neighbor coupling, we then have
\begin{eqnarray}
\frac{d\ln |G_{s(N-m+s)}|}{d\beta}=1
\end{eqnarray}
for each sub-chain, when $E_r$ is enclosed by the loop-like spectrum of each sub-chain.
This result indicates that each sub-chain has its own spectral winding number $\nu_s(E_r)=1$, but for the original 1D chain with $m$th-nearest neighbor couplings, the element $G_{1N}$ and $G_{N1}$ are zero as sites $1$ and $N$ belong to different decoupled sub-chains.

On the other hand, the $m$-sub-chain picture here also indicates an effective unit-cell structure with $m$ sublattices, even though the sublattices are physically equivalent on the lattice.
Thus the directional amplification of the overall system comprised by these sub-chains/sublattices shall be described by the combination of that of each sub-chain, corresponding to the corner blocks of the overall Green function matrix,
\onecolumngrid
\begin{eqnarray}
G_{\leftarrow,m\times m}=
\left(\begin{array}{cccc}
G_{1(N-m+1)} & G_{1(N-m+2)} &\cdots & G_{1N}\\
G_{2(N-m+1)} & G_{2(N-m+2)} &\cdots & G_{2N}\\
\vdots & \vdots  &\vdots  &\vdots \\
G_{m(N-m+1)} & G_{m(N-m+2)} &\cdots & G_{mN}\\
\end{array}\right),  \\
G_{\rightarrow,m\times m}=
\left(\begin{array}{cccc}
G_{(N-m+1)1} & G_{(N-m+1)2} &\cdots & G_{(N-m+1)m}\\
G_{(N-m+2)1} & G_{(N-m+2)2} &\cdots & G_{(N-m+2)m}\\
\vdots & \vdots  &\vdots  &\vdots \\
G_{N1} & G_{N2} &\cdots & G_{Nm}\\
\end{array}\right),
\end{eqnarray}
\twocolumngrid\noindent for signals moving toward the left- and the right side, respectively.
In the above schematic scenario where the sub-chains are fully decoupled from each other, only the diagonal elements of the above  two matrices are nonzero, i.e. 
$(G_{\leftarrow,m\times m})_{ab}\propto  \delta_{ab}e^{\beta}$ when $E_r$ is enclosed by the spectrum on the complex plane, and $(G_{\rightarrow,m\times m})_{ab}\propto  \delta_{ab}e^{-\beta}$ when $E_r$ is NOT enclosed by the spectrum on the complex plane.
This being the case, we arrive at
\begin{eqnarray}
\det[G_{\leftarrow,m\times m}]\propto e^{m\beta},
~\det[G_{\rightarrow,m\times m}]\propto e^{-m\beta},
\end{eqnarray}
in the above two cases repectively.
Thus we have 
 \begin{eqnarray}
\nu_{\leftarrow,m}:=\frac{d\ln |G_{\leftarrow,m\times m}|}{d\beta}=m,
\end{eqnarray}
corresponding to the spectral winding number $\nu(E_r)=m$ for the case with only the $m$th-nearest neighbor couplings.
Furthermore, this conclusion shall also be valid when $N/m$ is not an integer. In such cases, the system still possesses the $m$-sub-chain picture, only that the two ends of one sub-chain are connected to those of other sub-chains now. This conclusion is also verified by our numerical calculations.   As also mentioned in the main text, the physical analysis here is also valid if the system has many different types of coupling coexisting, but is still dominated by one type of coupling.  The topological robustness of the response can then retain the quantization.

\clearpage

\onecolumngrid
\begin{center}
\textbf{\large Supplementary Materials}\end{center}
\setcounter{equation}{0}
\setcounter{figure}{0}
\renewcommand{\theequation}{S\arabic{equation}}
\renewcommand{\thefigure}{S\arabic{figure}}
\renewcommand{\cite}[1]{\citep{#1}}
\section{Cases with couplings across different ranges}
In the main text, we have already witnessed an intriguing example where both nearest-neighboring and next-nearest-neighbor couplings present in the system.  In that case, with respect to different reference energy points, the spectral winding can have different nonzero values and this is also manifested as different quantized responses.  Benchmarking this  with our decoupled sub-chain picture, this indicates that there is competition in different regimes, and the underlying topological robustness associated with either single-chain physics or two coupled sub-chain physics
still yields, remarkably, quantized responses.  The next question is then, if we add more and more complexity to a non-Hermitian lattice system with many coexisting hopping length scales, can we still observe quantization  and hence a clear correspondence between spectral winding and the signal amplification.   Our answer is yes based on more computational tests.

\begin{figure*}[h]
\includegraphics[width=1\linewidth]{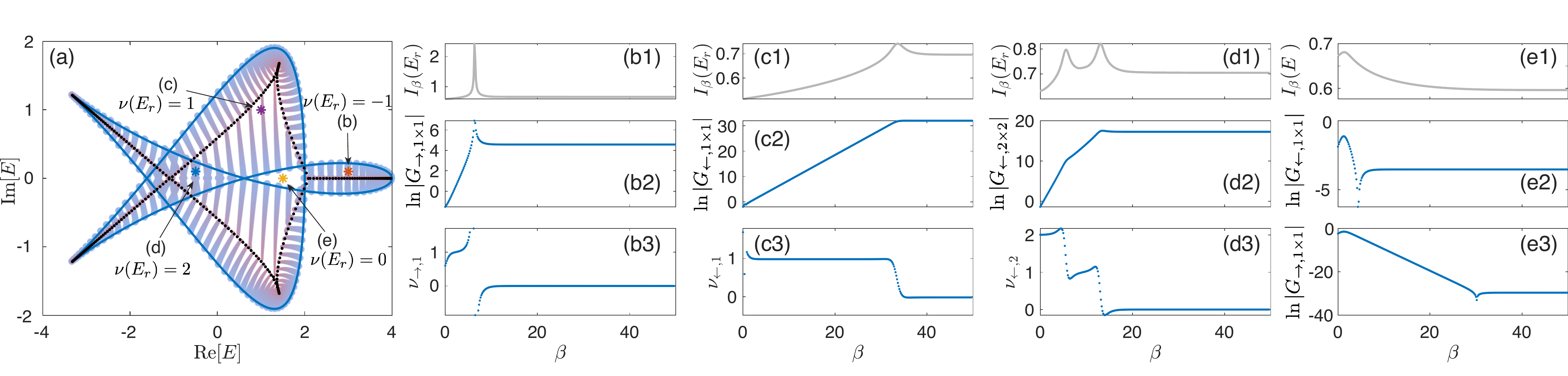}
\caption{(a) PBC (blue loop) and OBC (black dots) spectra of a system with many coexisting hopping length scales, and the PBC-OBC spectral evolution between them (blue-purple curves), tuned by the parameter $\beta$.  Four reference energy values $E_r$ are chosen to inspect the possible quantization of the gradient of the logarithm of directional amplification with respect to $\beta$, each with a different winding number $\nu(E_r)$, as indicated by the four colored marks.
(b)-(e) show the summed reverse energy spacing $I_\beta(E_r)=\sum_n |1/(E_n(\beta)-E_r)|$, logarithm of the amplification ratio for the block of $G$ associated with the winding number $\nu(E_r)$, as well as $\nu_{\leftarrow,m}$ and $\nu_{\rightarrow,m}$, as functions of $\beta$. In (e) we have $\nu(E_r)=0$, hence we only illustrate amplification ratios $G_{\leftarrow,1\times 1}$ and $G_{\rightarrow,1\times 1}$, which are both less than unity, indicating no amplification for a signal moving either toward the left or the right.
Parameters are $t_{2}=2$, $t_{-2}=1$, $t_3=0$, $t_{-3}=1$, and $t_{\pm m}=0$ for all other values of $m$, with $N=300$ lattice sites.
}
\label{fig:t2t3}
\end{figure*} 

Consider then a lattice model with nonzero terms
$\{t_{-r},t_{-r+1},...,t_{l-1},t_{l}\}$, and none of them is dominating over the rest.   We can still view this system
as one comprised of $m$ sub-chains, with $m\leqslant {\rm Max}[r,l]$.  We can investigate if the same response functions $\nu_{\leftarrow,m}$ or  
$\nu_{\rightarrow,m}$ can reflect the spectral winding behaviors.  To proceed specifically, consider the following Hamiltonian as an example,
\begin{eqnarray}
H=\sum_{x=1}^N\sum_{j=-r}^{l} t_j\hat{c}^\dagger_j\hat{c}_{x+j},
\end{eqnarray}
with $r=l=3$, and the boundary coupling tuned via  $t_j\rightarrow e^{-\beta}t_j$ when it connects sites at different ends of this 1D chain. 
The rather complicated spectral winding behavior is shown in Fig.~\ref{fig:t2t3}(a), indicating winding numbers -1, 1, 2, and 0.
As shown in Fig.~\ref{fig:t2t3}(b)-(d),
the obtained $\nu_{\leftarrow,m}$ or $\nu_{\rightarrow,m}$ for $m=1$ or $m=2$ still shows relatively clear plateaus for cases with nonzero $\nu(E_r)$, with the transitions of these plateaus in excellent agreement with the critical $\beta$ values for which the spectral winding numbers make jumps.   In panel (c3), the quantization in $\nu_{\leftarrow,1}$  is  clearly seen.  A careful reader might notice that in panel (b3) and (d3), the plateaus of the obtained response function $\nu_{\rightarrow,1}$ are so not clearly quantized.   To double check if this is merely a finite-size effect,  we
have increased the size of the model system and then much better quantized plateaus are indeed observed, as presented in Fig.~\ref{fig:size}.   
As to the case of $\nu(E_r)=0$ labeled by the yellow star in Fig.~\ref{fig:t2t3}(a), no amplification is obtained for a signal moving toward either the left- or the right-hand side, as indicated by the always-negative $\ln|G_{\leftarrow,1\times 1}|$ and $\ln|G_{\rightarrow,1\times 1}|$ in Fig.~\ref{fig:t2t3}(e). 
\begin{figure}[h]
\includegraphics[width=0.6\linewidth]{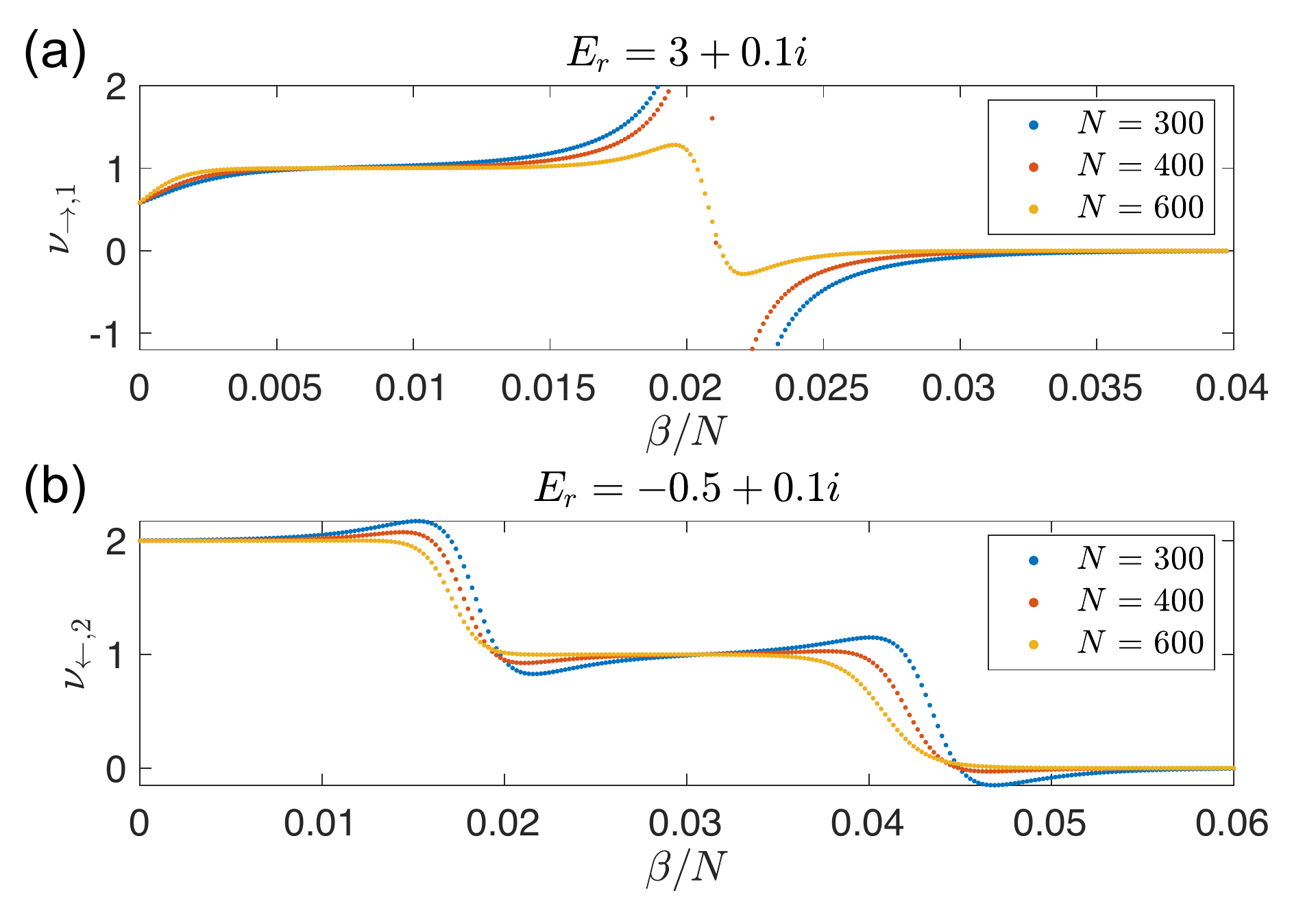}
\caption{(a) $\nu_{\rightarrow,1}$ and (b) $\nu_{\leftarrow,2}$ as functions of $\beta/N$ for the two cases in Fig.~\ref{fig:t2t3}(b) and (d) respectively, with different numbers of lattice $N=300$, $400$, and $600$. 
As our previous analytical results suggest that the transition value $\beta_c$ is proportional to $N$,
here the variable $\beta$ is rescaled by a factor of $1/N$, so as to map the transition points for systems with different sizes to the same parameter.
The plateaus are seen to be flatter (hence better quantization) for larger systems. 
Other parameters are $t_1=1$ and $t_{-1}=0.5$.
}
\label{fig:size}
\end{figure}

\section{Illustrative details of circuit for probing the response of the Hatano-Nelson model}

To be concrete, we provide explicit details for the realization of the simplest case of the circuit representing the Hatano-Nelson model with topological winding $\nu=1$. It consists of a chain of nodes connected by unbalanced couplings which simultaneously give rise to non-Hermiticity and non-reciprocity. It is well-established that such couplings can be realized with INICs (negative impedance converters with current inversion)~\cite{hofmann2019chiral,helbig2020generalized}, which contains operation amplifiers that break the reciprocity. 
To realize the tuning of the spectral reference point $\Omega$ and end-to-end couplings $e^{-\beta(\omega)}$, we also include tunable inductors and additional RLC elements as according to Ref.~\cite{li2020impurity}. Its Laplacian, together with these tunable elements, takes the form
\begin{align}
J=& \biggl[e^{-\beta(\omega)}\left(e^\alpha|N\rangle\langle 0|+e^{-\alpha}|0\rangle\langle N|\right)+\sum_{x=0,\pm}^{N-1}e^{\pm\alpha}|x\rangle\langle x\pm 1|
\notag\\
&-\sum_{x=0}^{N}\left(2 \cosh\alpha-\omega_0^2/\omega^2+\Omega(\omega)\right)|x\rangle\langle x|\biggl]\times i\omega C
\label{J}
\end{align}
where $\alpha=\tanh^{-1}\frac{C_1}{C_2}$, $C=\sqrt{C_2^2-C_1^2}$, $\omega_0^{-2}=l_{gr}C$ and $\Omega = \frac{-C_\Omega+iR_\Omega/\omega}{C}$. $C_1$ and $C_2$ are capacitors involved in internode couplings as according to Ref.~\cite{li2020impurity}, and $R_\Omega$, $l_{gr},C_\Omega$ are RLC elements that connect each node to the ground. By varying the choice of $C_\Omega$ and $R_\Omega$, $\Omega$ and hence different points of the complex eigenvalue plane can be sampled. In this circuit, the $\nu=1$ and $\nu=0$ regions are separated by the curve $\Omega=2\cosh(\alpha+ik)-2\cosh\alpha-\omega_0^2/\omega^2$. The effective flux $\beta(\omega)$ can be adjusted by tuning the AC frequency $\omega$, and is given by $\beta(\omega)=\ln[1-\omega^2l(C_2-C_1)]$, where $l$ is an inductor involved in the internode couplings which controls the sensitivity of the tuning. Clearly, by including additional couplings between more distant nodes~\cite{li2019emergence,lee2020imaging,ezawa2019electric2,lu2019probing,zhang2020topolectrical,song2020experimental,stegmaier2020topological}, this circuit construction can be extended to models with additional further couplings, such as that shown in the main text with $t_1=1,t_{-1}=0.5,t_2=2,t_{-2}=0$. 

%

\end{document}